# Comparative Investigation of the High Pressure Autoignition of the Butanol Isomers

*B.W. Weber and C.J. Sung*

*Department of Mechanical Engineering*
*University of Connecticut, Storrs, Connecticut 06269, USA*

Investigation of the autoignition delay of the butanol isomers has been performed at elevated pressures of 15 bar and 30 bar and low to intermediate temperatures of 680–860 K. The reactivity of the stoichiometric isomers of butanol, in terms of inverse ignition delay, was ranked as $n$-butanol > $sec$-butanol ~ $iso$-butanol > $tert$-butanol at a compressed pressure of 15 bar but changed to $n$-butanol > $tert$-butanol > $sec$-butanol > $iso$-butanol at 30 bar. For the temperature and pressure conditions in this study, no NTC or two-stage ignition behavior were observed. However, for both of the compressed pressures studied in this work, $tert$-butanol exhibited unique pre-ignition heat release characteristics. As such, $tert$-Butanol was further studied at two additional equivalence ratios ($\phi$ = 0.5 and 2.0) to help determine the cause of the heat release.

## 1. Introduction

One of the most common additives in gasoline is ethanol, a biofuel intended to reduce our dependence on petroleum as a source of energy. Many concerns have been raised, however, about whether ethanol is actually an improvement on the status quo [1]. Due to these issues, $n$-butanol has been suggested as a fuel that will alleviate many of the technical problems with ethanol [2]. Significant effort has been expended recently to determine the suitability of $n$-butanol for production in bio-processes. Although it is not clear which method is likely to produce butanol the most efficiently [3], research has typically focused on ways to improve $n$-butanol yield through genetic modification of typical industrial bacterial strains [4], as well as improvement of strains of bacteria that produce butanol naturally [5].

Despite the substantial work done to improve the bioproduction processes of $n$-butanol, until recently, little had been done to determine its suitability as a replacement for ethanol from a combustion perspective. However, the number of studies of $n$-butanol combustion has increased dramatically in the last few years. A small sampling of recent results includes flame speeds [6], ignition delays [7-11], and pyrolysis studies [12]. Although there are fewer studies of the isomers of $n$-butanol (i.e. $sec$-butanol, $tert$-butanol and $iso$-butanol), similar types of results are available (e.g. [13-15]). However, there is a scarcity of data at higher pressures and lower temperatures, especially for ignition delays. In this study, autoignition delay results collected using a heated Rapid Compression Machine (RCM) are presented for the four isomers of butanol at elevated pressure and low to intermediate temperature conditions.

## 2. Experimental

The Rapid Compression Machine (RCM) used in the current study has been described elsewhere [16]. The basic details are provided here for reference. The present RCM is a pneumatically-driven/hydraulically-stopped arrangement, which provides for compression times on the order of 30 ms. The states in the reaction chamber when the piston reaches Top Dead Center (TDC) are referred to as the compressed conditions. The initial temperature, initial pressure, and compression ratio can be varied to vary the compressed temperature ($T_C$) and compressed pressure ($P_C$) independently.

The procedure used in this work to create fuel/oxidizer premixtures has been described previously [8]. Fuel/oxidizer premixtures were made in a 17 L mixing tank, equipped with heaters and a magnetic stirring apparatus. The reaction chamber of the RCM was also heated, allowing the entire system to reach temperatures up to 140 °C. This allows fuels with rather low vapor pressure to be studied in the RCM. The preheat temperature of the mixing tank was set above the saturation temperature of the fuels to ensure their complete vaporization. The saturation vapor dependence of the fuels was taken from the *Chemical Properties Handbook* by Yaws [17].

Experiments were carried out at the same pressure and equivalence ratio condition for all four isomers of butanol. One set of experiments was carried out at $P_C$=15 and 30 bar, for $\phi$=1.0 mixture in synthetic nitrogen-oxygen air. The corresponding reactant mole fractions were: $X_{fuel} = 0.0338$, $X_{O2} = 0.2030$, and $X_{N2} = 0.7632$. Two additional data sets at $\phi$=0.5 and 2.0 in air and $P_C = 30$ bar were taken for *tert*-butanol only. Overall, the data sets spanned the compressed temperature ($T_C$) range from 680 K to 860 K.

The end of compression, when the piston reached TDC, was identified by the maximum of the pressure trace ($P(t)$) prior to the ignition point. The local maximum of the derivative of the pressure trace with respect to time ($P'(t)$), in the time after TDC, was defined as the point of ignition. The ignition delay was the time difference between the point of ignition and the end of compression. Figure 1 illustrates the definition of ignition delay ($\tau$) used in this study.

Temperature at TDC was used as the reference temperature for reporting ignition delay data and was called the compressed temperature ($T_C$). The temperature was calculated using the adiabatic core assumption by matching a simulated compression stroke to the actual pressure trace. A non-reactive experiment is performed to ensure that no significant heat release occurs during the compression stroke. The non-reactive run is set up by replacing the oxygen in the mixture with nitrogen, to eliminate oxidation reactions but maintain a similar heat capacity ratio. An example non-reactive pressure trace can be seen in Figure 1. Further details and validation of this approach can be found in Ref. [16].

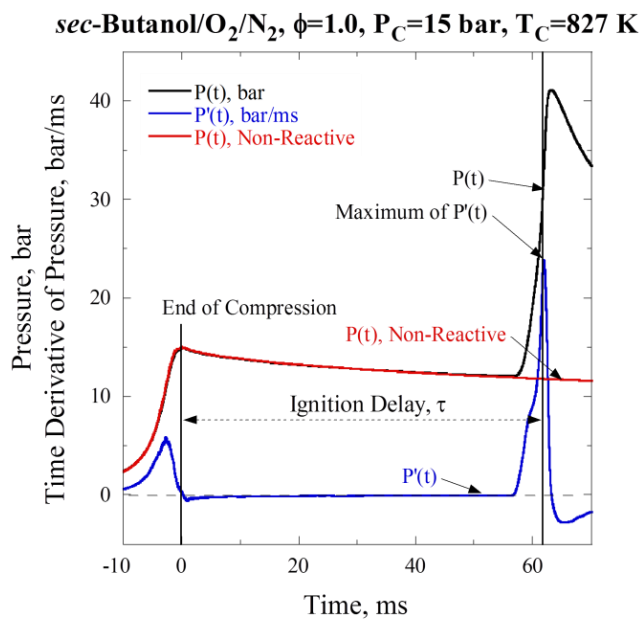

**Figure 1. Definition of ignition delay used in this study. $P'(t)$ is the time derivative of the pressure profile.**

### 3. Results and Discussion

Figure 2 shows Arrhenius plots of ignition delays of the four isomers of butanol. Figure 2(a) shows the ignition delays at $P_C = 15$ bar, while Figure 2(b) shows the ignition delays at $P_C = 30$ bar. The error bars on these figures represent two standard deviations calculated from the experiments, while the dashed lines are least squares fits to the data. The difference in order of reactivity is quite clear from these graphs – at 15 bar, the reactivities are in the order $n$-butanol $> sec$-butanol $\sim iso$-butanol $> tert$-butanol, while at 30 bar, the reactivities are in the order $n$-butanol $> t$-butanol $> sec$-butanol $> iso$-butanol. The results at 15 bar generally agree with results found previously by Moss et al. [14] and Veloo and Egolfopoulos [13], although it appears as temperature decreases that $tert$-butanol may become more reactive than $iso$- or $sec$-butanol. Due to limitations of the RCM, lower temperature comparison for 15 bar cases were unable to be obtained; however, increasing the compressed pressure to 30 bar showed that under the right conditions, $tert$-butanol becomes more reactive than both $iso$- and $sec$-butanol.

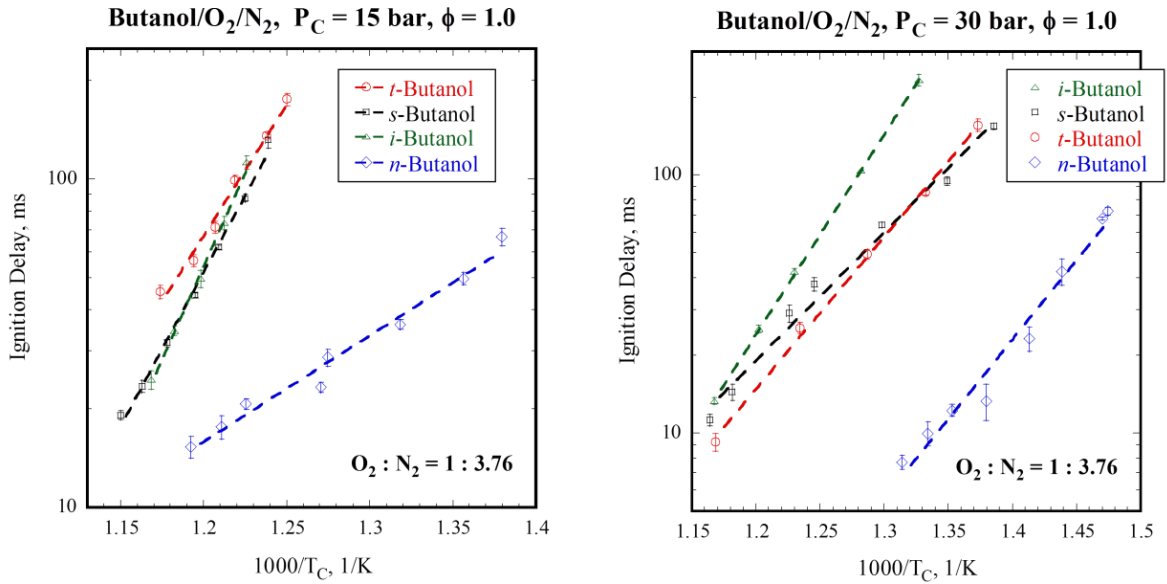

**Figure 2.** (a) Arrhenius plot of the ignition delay of the four isomers of butanol at compressed pressure $P_C = 15$ bar. (b) Arrhenius plot of the ignition delay of the four isomers of butanol at compressed pressure $P_C = 30$ bar.

In addition to becoming relatively more reactive, $tert$-butanol shows very interesting pre-ignition heat release effects. This effect is demonstrated in Figure 3, which shows the pressure traces of RCM experiments for $tert$-butanol at $\phi = 1.0$ and $P_c = 30$ bar. Very clear deviation from the non-reactive pressure trace, shown in grey, is evident, indicating substantial pre-ignition heat release.

To help determine what is causing this pre-ignition behavior, which is not predicted by chemical kinetic mechanisms available in the literature, experiments at two additional equivalence ratios were conducted. Figure 4 is an Arrhenius plot of the ignition delays of $tert$-butanol, at equivalence ratios of $\phi = 0.5, 1.0,$ and $2.0$ for $P_C = 30$ bar. Once again, error bars are two standard deviations of the experimental data and the dashed lines are least squares fits to the data. As expected, the lower equivalence ratio is the least reactive, and the highest equivalence ratio is the most reactive, under the conditions investigated. In addition, the overall activation energy does not appear to change – that is, all the lines have similar slopes. Further analyses, both computational and experimental, are required to determine the cause of this interesting pre-ignition heat release behavior.

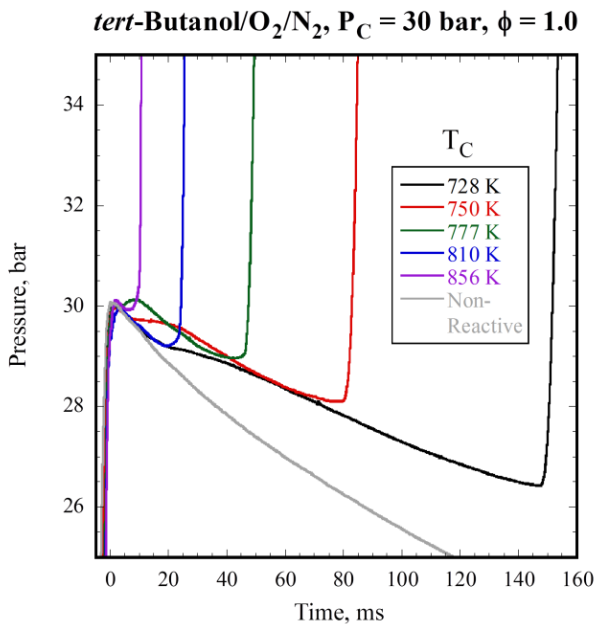

**Figure 3.** Plot of the pressure traces from the RCM for $tert$-butanol at $\phi = 1.0$ and $P_c = 30$ bar

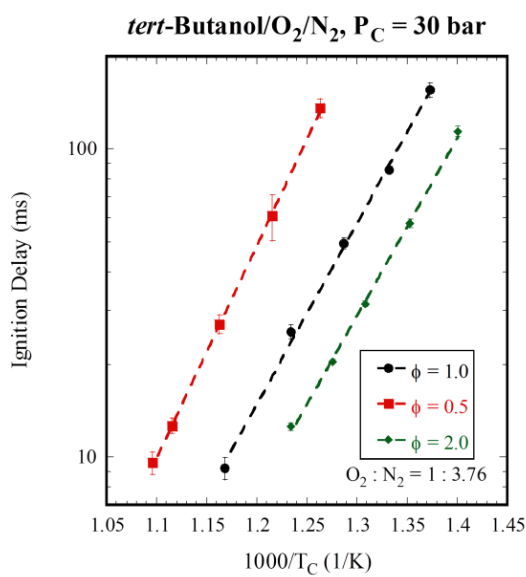

**Figure 4.** Arrhenius plot of the ignition delays of $tert$-butanol at three equivalence ratios, $\phi = 0.5, 1.0,$ and $2.0$ and compressed pressure $P_C = 30$ bar.

## Acknowledgements

This material is based upon work supported as part of the Combustion Energy Frontier Research Center, an Energy Frontier Research Center funded by the U.S. Department of Energy, Office of Science, Office of Basic Energy Sciences, under Award Number DE-SC0001198.